\documentclass[twocolumn,english,aps,superscriptaddress,prb]{revtex4-2}

\usepackage[colorlinks=true,allcolors=blue]{hyperref}
\usepackage{array}
\usepackage[latin9]{inputenc}
\usepackage{amsmath}
\usepackage{amssymb}
\usepackage{graphicx}
\usepackage{babel}
\usepackage{mathrsfs}
\usepackage{amsfonts}
\usepackage{epstopdf}
\usepackage{multirow}
\usepackage{color}
\usepackage{natbib}
\usepackage{bm}
\usepackage{soul}

\usepackage{xcolor}

\begin{document}

\title{Quantized nonlinear transport and its breakdown in Fermi gases with Berry curvature}
\author{Fan Yang}
\email{101013867@seu.edu.cn}
\affiliation{Key Laboratory of Quantum Materials and Devices of Ministry of Education, School of Physics, Southeast University, No.2 SEU Road, Nanjing, China 211189}

\author{Xingyu Li}
\email{ xyli22@mails.tsinghua.edu.cn}
\affiliation{Institute for Advanced Study, Tsinghua University, Beijing, China}
\date{\today}
\begin{abstract}
    Quantized transport not only exist in gapped topological states but also in metallic states. Recently, Kane proposed a quantized nonlinear conductance in ballistic metals whose value is determined by the Euler characteristic of the Fermi sea [Phys. Rev. Lett. 128, 076801 (2022)]. In this paper, we consider two-dimensional noninteracting fermionic systems whose Fermi surface has nonvanishing Berry curvature. We find that the Berry curvature at the Fermi surface does not affect the quantized nonlinear transport for translationally invariant systems. When spatial inhomogeneity is introduced, such quantization breaks down due to the combined effect of Berry curvature and the gradient of local potential. Such breakdown of quantization can be observed in trapped ultracold atoms with topological bands. 
\end{abstract}

\maketitle

\section{Introduction}
Quantized transport is the hallmark of nontrivial topology in physical systems. The most well-known type of quantized transport is the quantized Hall and spin Hall conductance in insulators \cite{Klitzing80,Thouless82,Kane05b,Bernevig06a,Konig07}. Such quantized conductance reflects the topology of wavefunctions in momentum space. On the other hand, quantized conductance also exists in ballistic metals. This includes the quantized Landauer conductance in 1D nanowires \cite{Landauer57,vanWees88,Wharam88,Honda95,vanWeperen13,Frank98} and the recently proposed quantized nonlinear conductance in 2D electron gas \cite{Kane22}, both of which can also be observed in trapped atomic gases \cite{Krinner15,Krinner17,Lebrat19,Yang22QNL,PFZhang23}. Such quantized conductance in metals reflects a different type of topological invariants---the Euler characteristic of the Fermi sea.

The Euler characteristic is an important topological invariant of Fermi gases and Fermi liquids. It also leads to quantized rectified conductance of Andreev bound states in Josephson junctions \cite{Tam23,tam2023topological}, dictates multipartite entanglement entropy \cite{Tam22} and multi-point density correlation function \cite{Tam24}, and even constrains the value of topological invariants when metals become superconducting \cite{EulerChern,Jia24}.

Previous research on the quantized transport given by Euler characteristics has assumed partially filled bands without Berry curvature \cite{Kane22,Yang22QNL,PFZhang23,Tam23,tam2023topological}. Nontrival Berry curvature at the Fermi surface leads to anomalous Hall effect \cite{Haldane04,Sodemann15,XLChen22}, which is non-quantized and may enter the nonlinear conductance due to the geometry of the three-terminal setup for measurements (see Fig. \ref{illustration}a). In this paper, we examine the interplay between anomalous Hall effect and nonlinear transport in two-dimensional (2D) metals without interaction. We show that for spatially uniform metals, the anomalous Hall effect does not affect the quantization of nonlinear transport. However, for an inhomogeneous Fermi gas in an external potential varying slowly in space, such as cold atoms in a trap, this quantization breaks down when the Berry curvature is nonvanishing at the Fermi surface.

\section{Quantized nonlinear transport with Berry curvature}
\begin{figure}
    \centering
    \includegraphics[width=\columnwidth]{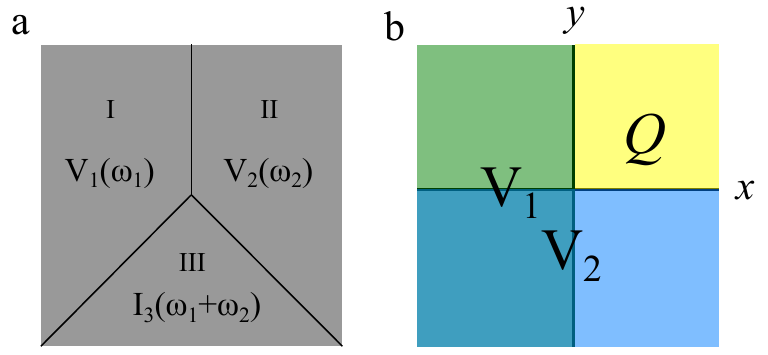}
    \caption{a: Illustration of three-terminal second-order nonlinear AC conductance measurement. The three terminals intersect at a single point and the frequency-dependent conductance is proportional to the Euler characteristic of the Fermi sea. b: This measurement is equivalent to a thought experiment with consecutive voltage pulses $V_1$ and $V_2$ applied in the left- and lower-half planes, respectively, which leads to quantized charge transport $Q$ to the first quadrant given by the Euler characteristic.}
    \label{illustration}
\end{figure}

The quantized nonlinear conductance in 2D exists in a three-terminal setting as shown in Fig. \ref{illustration}a. AC voltages $V_1(\omega_1)$ and $V_2(\omega_2)$ of frquencies $\omega_1$ and $\omega_2$ are applied in regions I and II, respectively, and the current $I_3(\omega_1+\omega_2)$ of frequency $\omega_1+\omega_2$ is measured in region III. If we assume no Berry curvature, the frequency-dependent nonlinear conductance is given by the Euler characteristic of the Fermi sea as shown in Ref. \cite{Kane22},
\begin{equation}
    G(\omega_1,\omega_2)=\frac{V_1(\omega_1)V_2(\omega_2)}{I_3(\omega_1+\omega_2)}=\frac{\omega_1+\omega_2}{i\omega_1\omega_2}\frac{e^3}{h^2}\chi_F,
\end{equation}
where $e$ is the electron charge, $h$ is the Planck constant, and $\chi_F$ is the Euler characteristic of the Fermi sea. In 2D, $\chi_F$ equals to the number of electron-like Fermi surfaces minus the number of hole-like Fermi surfaces (open Fermi surfaces do not contribute to $\chi_F$).

The quantized nonlinear transport can also be observed in a thought experiment shown in Fig. \ref{illustration}b \cite{Kane22,Yang22QNL}. Voltage pulses $V_1=\xi_1(h/e)\delta(t-t_1)\theta(-x)$ and $V_2=\xi_2(h/e)\delta(t-t_2)\theta(-y)$ are applied in the left and lower half planes at time $t_1$ and $t_2$, respectively. $\theta(\cdot)$ is the Heaviside step function. At a later time $t_3>t_2>t_1$, the excess charge transported to the first quadrant cooperatively by the two pulses is quantized and given by $Q=e\xi_1\xi_2\chi_F$. That is we measure the extra charges in the first quadrant after the two pulses subtracted by the extra charges transferred by each pulse individually. See Ref. \cite{Yang22QNL} for a detailed description of the experimental protocol.

In order to observe the quantized transport, it is crucial that regions I, II, and III intersect at a single point with all the angles formed by the boundaries less than $\pi$ \cite{Kane22}. Under such geometry, when transverse transport exists on the Fermi surface, it is not immediately clear whether the quantized conductance would be affected. In the following, we will consider metals with nonvanishing Berry curvature and analyze whether the anomalous Hall effect would affect nonlinear transport. 
We focus on a slightly modified version of the thought experiment. 
Instead of measuring $Q$ in the first quadrant, we measure it in the region $\Sigma=[0,L]\times[0,L]$ with $L>v_\text{0,max}t_{31}$, where  $t_{ij}=t_i-t_j$ and $v_\text{0,max}=\text{max}\{|{\bf v_0}|\}$ with ${\bf v_0}$ being the velocity of electrons in the Fermi sea in the absence of applied pulses. The reason for measuring $Q$ in a bounded area will become clear shortly.  

The excess charge transported to region $\Sigma$ can be computed using semiclassical methods \cite{Xiao10}.
The semiclassical equations of motion for electron wavepackets are
\begin{equation}
    \dot{\bf k}=\frac{e}{\hbar}{\bf E}+\frac{e}{\hbar}\dot{\bf r}\times{\bf B},
\end{equation}
\begin{equation}
    \dot{\bf r}=\frac{1}{\hbar}\nabla_{\bf k}\epsilon_{\bf k}+{\dot{\bf k}}\times{\bm\Omega}_{\bf k},
\end{equation}    
where ${\bf E}$ and ${\bf B}$ are electric and magnetic fields, respectively, $\epsilon_{\bf k}$ is the energy dispersion of the partially filled band, and ${\bm \Omega}_{\bf k}=\Omega_{\bf k}\hat{z}$ is the Berry curvature. For electric conductance measurements, we have ${\bf B}=0$. In previous studies, it was implicitly assumed $\Omega_{\bf k}=0$. In the following, we will consider a partially filled band with $\Omega_{\bf k}\neq0$. We note that the partially filled band can be either topological or trivial, but its Berry curvature on the Fermi surface does not vanish.

Let us consider noninteracting electrons. The phase space distribution function $f({\bf r},{\bf k},t)$ follows the collisonless Boltzmann equation
\begin{equation}\label{boltzmann}
    \partial_tf+{\bf v}\cdot\nabla_{\bf r}f+\frac{1}{\hbar}{\bf F}\cdot\nabla_{\bf k}f=0,
\end{equation}
where ${\bf v}=(1/\hbar)\nabla_{\bf k}\epsilon_{\bf k}+(1/\hbar){\bf F}\times{\bm \Omega}_{\bf k}$ is the velocity of the wavepacket, and ${\bf F}=-e\nabla_{\bf r}(V_1+V_2)={\bf F}_1+{\bf F}_2$ with
\begin{eqnarray}
    &{\bf F}_1=\xi_1h\delta(t-t_1)\delta(x)\hat{x},\label{F1}\\
    &{\bf F}_2=\xi_2h\delta(t-t_2)\delta(y)\hat{y}.\label{F2}
\end{eqnarray}

Let $f_0({\bf r},{\bf k},t)$ be the distribution function at equilibrium, which satisfies
\begin{equation}
    \partial_t f_0+{\bf v_0}\cdot\nabla_{\bf r} f_0=0,
\end{equation}
with ${\bf v_0}=(1/\hbar)\nabla_{\bf k}\epsilon_{\bf k}$. The Liouville's theorem guarantees $f_0({\bf r},{\bf k},t)=f_0({\bf r}(t),{\bf k}(t))$, where ${\bf r}(t)$ and ${\bf k}(t)$ are given by the trajectory of wavepackets in the phase space. In the absence of applied pulses, we have
\begin{eqnarray}\label{ch1}
    {\bf r}(t)=&&{\bf r}+{\bf v_0}t,\\
    {\bf k}(t)=&&{\bf k}.    \label{ch2}
\end{eqnarray}
Following the procedures outlined in Refs. \cite{Kane22,Yang22QNL}, the Boltzmann equation can be solved by the method of characteristics \cite{MathPhys} with the characteristic lines given by Eqs. (\ref{ch1}) and (\ref{ch2}). The change of distribution function due to the combined effect of both voltage pulses is
\begin{multline}
    \delta f_{12}=-(2\pi)^2\xi_1\xi_2\delta(y_2)\left(\Omega_{\bf k}\frac{\partial}{\partial x_2}+\frac{\partial}{\partial k_y}\right)\\
    \times\left[\delta(x_1)\left(\Omega_{\bf k}\frac{\partial}{\partial y_1}-\frac{\partial}{\partial k_x}\right)\right]f_0,
\end{multline}
where $f_0({\bf r}(t),{\bf k}(t))=\theta(\epsilon_F-\epsilon_{\bf k})$, $\epsilon_F$ is the Fermi energy, and $x_i=x(t_i)$ and $y_i=y(t_i)$ are given by Eq. (\ref{ch1}).

The excess charge transported cooperatively by the two pulses to the region $\Sigma$ is
\begin{equation}
    Q=e\int\frac{d^2{\bf k}}{(2\pi)^2}\int_\Sigma d^2{\bf r}_3 \delta f_{12}.
\end{equation}
The spatial integration can be performed easily by utilizing $L>v_{0,x}t_{21}, v_{0,y}t_{31}$.
Noting that $\partial f_0/\partial y_1=0$, we can split $Q$ into two parts $Q=Q_0+Q'$, with
\begin{eqnarray}
    Q_0=e\xi_1\xi_2\int{d^2{\bf k}}\theta(v_{0,y})\frac{\partial}{\partial k_y}\left[\theta(v_{0,x})\frac{\partial f_0}{\partial k_x}\right],
\end{eqnarray}
and 
\begin{equation}
    Q'=-e\xi_1\xi_2\int{d^2{\bf k}}\theta(v_{0,y})\theta(v_{0,x})\delta(v_{0,x}t_{31})\Omega_{\bf k}\frac{\partial f_0}{\partial k_x}.
\end{equation}
$Q_0$ is exactly the charge transported in a trivial band without Berry curvature, which was shown in Ref. \cite{Kane22} to be given by the Euler characteristic $Q_0=e\xi_1\xi_2\chi_F$.
$Q'$ is the Berry curvature dependent part. It is easy to show that $Q'$ vanishes. As we have $\partial f_0/\partial k_x=\hbar v_{0,x}\partial f_0/\partial \epsilon_{\bf k}$, the integrand contains a factor of $v_{0,x}\delta(v_{0,x}t_{31})$, resulting in $Q'=0$.

The reason we are interested in the bounded area $\Sigma$ is as follows. When the voltage pulses are applied, the wavepackets at the boundary of the pulses gain an instantaneous anomalous velocity and are ``pumped" {\it along} the boundary. {This ``pump effect" does not change the distribution function $f$. Therefore, it cannot be captured by the previous calculations, which compute the excess charge through the change of $f$.} However, the pump effect leads to a finite \textit{instantaneous} current flowing into or out of the first quadrant. By selecting a bounded area $\Sigma$ and only measuring the difference in total charge, this ``pump effect" is canceled between opposite boundaries of $\Sigma$. In this way, we only measure the effect of sustained current after the two pulses.

In conclusion, for 2D noninteracting electron gas with translation symmetry, Berry curvature on the Fermi surface does not affect the quantized nonlinear transport as long as one measures the charge difference in a bounded area. The key reason is that the nonlinear transport probes electron properties at equilibrium where anomalous velocities vanish. Thus, the nonlinear conductance is solely determined by the energy dispersion.

\section{Spatial inhomogeneity and the breakdown of quantization}

Without Berry curvature, the quantized nonlinear transport can also be observed in trapped atomic gases. If the trap potential varies slowly in space such that Fermi surfaces are well-defined locally, the quantized nonlinear transport is given by the Euler characteristic of the local Fermi sea at the intersection of the two pulses \cite{Yang22QNL}.

In the following, we will show that such quantized transport breaks down due to the combined effect of Berry curvature and spatial inhomogeneity. Let us consider Fermi gases with Berry curvature in an external trap potential $U_{\bf r}$. At equilibrium, the distribution function becomes $f_0({\bf r}(t),{\bf k}(t))=\theta(\epsilon_F-\epsilon_{\bf k}-U_{\bf r})$ and the equations of motion in the absence of the applied pulses are
\begin{equation}\label{eq:k}
    \dot{\bf k}=-\frac{1}\hbar\nabla_{\bf r}U_{\bf r},
\end{equation}
\begin{equation}\label{eq:r}
    \dot{\bf r}=\frac{1}{\hbar}\nabla_{\bf k}\epsilon_{\bf k}-\frac{1}{\hbar}\nabla_{\bf r}U_{\bf r}\times{\bm \Omega}_{\bf k}.
\end{equation}
Here, $\epsilon_{\bf k}$ is the energy dispersion of the partially filled band without the external trap. $E=\epsilon_{{\bf k}(t)}+U_{{\bf r}(t)}$ is conserved, i.e., $dE/dt=0$.

%In this case, the momentum is not conserved. Therefore, strictly speaking, we are no longer in the ballistic transport regime. However, since external traps are essential for cold atom experiments, it is still crucial to understand its effect on transport. Let us consider slowly varying trap  such that Fermi surfaces are well-defined locally in space. 

Accordingly, in the Boltzmann equation (\ref{boltzmann}), ${\bf v}$ and ${\bf F}$ are given by
\begin{equation}
   {\bf v}=\frac{1}{\hbar}\nabla_{\bf k}\epsilon_{\bf k}-\frac{1}{\hbar}\nabla_{\bf r}U_{\bf r}\times{\bm \Omega}_{\bf k}+\frac1\hbar ({\bf F}_1+{\bf F}_2)\times{\bm \Omega}_{\bf k}, 
\end{equation}
\begin{equation}
    {\bf F}=-\frac1\hbar\nabla_{\bf r}U_{\bf r}+{\bf F}_1+{\bf F}_2,
\end{equation}
with ${\bf F}_1$ and ${\bf F}_2$ given by Eqs. (\ref{F1},\ref{F2}).
The distribution function at equilibrium now satisfies
\begin{equation}
    \partial_t f_0+{\bf v_0}\cdot\nabla_{\bf r} f_0-\frac1\hbar\nabla_{\bf r}U_{\bf r}\cdot\nabla_{\bf k}f_0=0,
\end{equation}
with $\bf v_0=\frac{1}{\hbar}\nabla_{\bf k}\epsilon_{\bf k}-\frac{1}{\hbar}\nabla_{\bf r}U_{\bf r}\times{\bm \Omega}_{\bf k}$.

Generally speaking, we do not have an analytical solution to the equations of motion (\ref{eq:k}) and (\ref{eq:r}). However, the solution to the Boltzmann equation can still be formally written as
\begin{multline}
    \delta f_{12}=-(2\pi)^2\xi_1\xi_2\delta(y_2)\left(\Omega_{\bf k}\frac{\partial}{\partial x_2}+\frac{\partial}{\partial k_{y2}}\right)\\
    \times\left[\delta(x_1)\left(\Omega_{\bf k}\frac{\partial}{\partial y_1}-\frac{\partial}{\partial k_{x1}}\right)\right]f_0,
\end{multline}
where $k_{xi}=k_x(t_i)$, $k_{yi}=k_y(t_i)$ and $x_i=x(t_i)$, $y_i=y(t_i)$ follow the equations of motion (\ref{eq:k},\ref{eq:r}).
Following Ref. \cite{Yang22QNL}, we consider the short time limit $t_{31}\to0$, in which case we only probe the Fermi surface defined locally at the intersection of the two pulses. In this limit, we can linearize the trajectroy of the wavepackets inside the $\delta$-functions
\begin{equation}
    {\bf r}_i\approx{\bf r}_3-{\bf v_0}t_{3i},
\end{equation}
and use ${\bf k}_i\approx{\bf k}$ and ${\bf r}_i\approx{\bf r}$ when taking derivatives.
Notice that the key difference here is that $\bf v_0=\frac{1}{\hbar}\nabla_{\bf k}\epsilon_{\bf k}-\frac{1}{\hbar}\nabla_{\bf r}U_{\bf r}\times{\bm \Omega}_{\bf k}$ now contains the anomalous part even at the equilibrium. In this case, the velocity ${\bf v_0}$ is no longer given by the gradient of the energy dispersion $\epsilon_{\bf k}$, breaking an important condition in linking the nonlinear response to Fermi sea topology with Morse theory \cite{Kane22}.

In the short time limit, the excess number of particles in region $\Sigma$ is 
\begin{multline}
    N=-\xi_1\xi_2\int d^2{\bf k}\int_\Sigma d^2{\bf r}_3\delta(y_3-v_{0,y}t_{32})\left(\Omega_{\bf k}\frac{\partial}{\partial x}+\frac{\partial}{\partial k_{y}}\right)\\
    \times\left[\delta(x_3-v_{0,x}t_{31})\left(\Omega_{\bf k}\frac{\partial}{\partial y}-\frac{\partial}{\partial k_{x}}\right)\right]f_0,
\end{multline}
It can be split into four parts by selecting one derivative from each bracket. In the following, we analyze them separately.

First, we have a term of the same form as the homogeneous case,
\begin{align}\label{N1}
    N_1&=\xi_1\xi_2\int d^2{\bf k}\int_\Sigma d^2{\bf r}_3\delta(y_2)\frac{\partial}{\partial k_y}\left[\delta(x_1)\frac{\partial}{\partial k_x}\right]f_0\nonumber\\
    &=\xi_1\xi_2\int d^2{\bf k}\theta(v_{0,y})\frac{\partial}{\partial k_{y}}\left[\theta(v_{0,x})\frac{\partial}{\partial k_{x}}\right]f_0\Bigg |_{\bf r=0}.
\end{align}
To write this equation in a form in resemblance of the Euler characteristic, we perform an integration by parts and add to the integrand a term
\begin{align}
    &\theta(v_{0,x})\frac{\partial\theta(v_{0,y})}{\partial k_x}\left(\frac{\partial f_0}{\partial k_y}+\Omega_{\bf k}\frac{\partial f_0}{\partial x}\right)\Bigg|_{\bf r=0}\label{null}\\
    =&\theta(v_{0,x})\frac{\partial v_{0,y}}{\partial k_x}\frac{d f_0}{d E}\hbar v_{0,y}\delta(v_{0,y})\Bigg|_{\bf r=0}.
\end{align}
Notice that $v_{0,y}\delta(v_{0,y})=0$, and thus, adding the above term do not alter the value of $N_1$. Incorporating Eq. (\ref{null}) into Eq. (\ref{N1}), we arrive at the following expression 
\begin{align}\label{extra}
    N_1=&\xi_1\xi_2\sum_{\bf v_0=0}\text{sgn}\det\left[\frac{\partial(v_{0,x},v_{0,y})}{\partial(k_x,k_y)}\right]\Bigg|_{\bf r=0}\nonumber\\
    &+\xi_1\xi_2\int d^2{\bf k}\Omega_{\bf k}\theta(v_{0,x})\frac{\partial\theta(v_{0,y})}{\partial k_x}\frac{\partial f_0}{\partial x}\Bigg|_{\bf r=0},
\end{align}
where $\partial(v_{0,x},v_{0,y})/\partial(k_x,k_y)$ is the Jacobian matrix, $\det[\cdot]$ is the determinant and the summation is for $\epsilon_{\bf k}+U_{\bf r=0}\leq\epsilon_F$. In the absence of either Berry curvature or inhomogeneity, the second term in Eq. (\ref{extra}) vanishes and the first term can be recast as the Morse theory expression for the Euler characteristic via ${\bf v_0}=(1/\hbar)\nabla_{\bf k}\epsilon_{\bf k}$ \cite{Kane22,Yang22QNL}.
However, ${\bf v_0}$ now contains the anomalous part ${\bf v}_a=-(1/\hbar)\nabla_{\bf r}U_{\bf r}\times{\bm \Omega}_{\bf k}$. Therefore, we can no longer rewrite the first term in Eq. (\ref{extra}) in terms of the derivatives of the dispersion $\epsilon_{\bf k}$, which is the key to connect the particle number to Euler characteristic of the Fermi sea
\begin{equation}
    \chi_F=\sum_{\nabla_{\bf k}\epsilon_{\bf k}=0}\text{sgn}\det \left[\frac{\partial^2\epsilon_{\bf k}}{\partial k_\mu\partial k_\nu}\right], 
\end{equation}
with $\mu,\nu$ standing for the subscripts $x,y$.
As a result, $N_1$ is determined not only by the Fermi sea, but also by the Berry curvature. 

The second part of $N$,
\begin{align}
    N_2&=-\xi_1\xi_2\int d^2{\bf k}\int_\Sigma d^2{\bf r}_3\delta(y_2)\frac{\partial}{\partial k_y}\left[\delta(x_1)\Omega_{\bf k}\frac{\partial}{\partial y}\right]f_0\nonumber\\
    &=\xi_1\xi_2\int d^2{\bf k}\Omega_{\bf k}\theta(v_{0,x})\frac{\partial\theta(v_{0,y})}{\partial k_y}\frac{\partial f_0}{\partial y}\Bigg|_{\bf r=0},
\end{align}
is closely related to the second term of $N_1$ in Eq. (\ref{extra}). We will examine them together later.

The other two parts can be shown to cancel each other.
The third and fourth parts are
\begin{align}
    N_3&=-\xi_1\xi_2\int d^2{\bf k}\int_\Sigma d^2{\bf r}_3 \delta(y_2)\Omega_{\bf k}\frac{\partial}{\partial x}\left[\delta(x_1)\Omega_{\bf k}\frac{\partial}{\partial y}\right]f_0\nonumber\\
    &=\xi_1\xi_2\int d^2{\bf k}\Omega_{\bf k}^2\theta(v_{0,y})\delta(v_{0,x} t_{31})\frac{\partial f_0}{\partial y}\Bigg|_{\bf r=0},
\end{align}
and
\begin{align}
    N_4&=\xi_1\xi_2 \int d^2{\bf k}\int_\Sigma d^2{\bf r}_3\delta(y_2)\Omega_{\bf k}\frac{\partial}{\partial x}\left[\delta(x_1)\frac{\partial}{\partial k_x}\right]f_0\nonumber\\
    &=-\xi_1\xi_2\int d^2{\bf k}\Omega_{\bf k}\theta(v_{0,y})\delta(v_{0,x} t_{31})\frac{\partial f_0}{\partial k_x}\Bigg|_{\bf r=0}.
\end{align}
By utilizing the relation 
\begin{align}
    \Omega_{\bf k}\frac{\partial f_0}{\partial y}-\frac{\partial f_0}{\partial k_x}=\frac{df_0}{dE}\left(\Omega_{\bf k}\frac{\partial U_{\bf r}}{\partial y}-\frac{\partial\epsilon_{\bf k}}{\partial k_x}\right)=\frac{df_0}{dE}\hbar v_{0,x}
\end{align}
and $v_{0,x}\delta(v_{0,x}t_{31})=0$, we have $N_3+N_4=0$.

By adding the four parts together, we arrive at the final expression for $N$
\begin{align}
    N=&\xi_1\xi_2\sum_{\bf v_0=0}\text{sgn}\det\left[\frac{\partial(v_{0,x},v_{0,y})}{\partial(k_x,k_y)}\right]\Bigg|_{\bf r=0}\nonumber\\
    &+\xi_1\xi_2\int d^2{\bf k} \Omega_{\bf k}\theta(v_{0,x})(\nabla_{\bf k } \theta(v_{0,y}))\cdot(\nabla_{\bf r} f_0)\Big|_{\bf r=0}.
\end{align}
The first term is still quantized. However, since the velocity ${\bf v_0}$ can no longer be written as a gradient field, we can no longer interpret it with Morse theory. Apart from the factor $\xi_1\xi_2$, the first term gives the sum of the indices of the zeros of the ${\bf v_0}$ field inside the local Fermi sea $\epsilon_{\bf k}+U_{\bf r=0}\leq\epsilon_F$. Therefore, as long as ${\bf v_0}$ is pointing outwards against the local Fermi surface, by Poincare-Hopf theorem \cite{TopoDiff}, the summation still gives the Euler characteristic of the local Fermi sea. Only when the anomalous velocity is large enough such that ${\bf v_0}$ no longer points outwards, its value changes.
The second term is not quantized.  $\nabla_{\bf r} f_0=-\delta(\epsilon_F-\epsilon_{\bf k}-U_{\bf r})\nabla_{\bf r}U$ singles out local Fermi surface. $\theta(v_{0,x})\nabla_{\bf k}\theta(v_{0,y})=\theta(v_{0,x})\delta(v_{0,y})\nabla_{\bf k}v_{0,y}$ singles out $v_{0,x}>0$, $v_{0,y}=0$. Therefore, only points on the local Fermi surface at the intersections of the two pulses with $v_{0,x}>0$ and $v_{0,y}=0$ contribute to the non-quantized part. But since the velocity now contains the anomalous part, $v_{0,x}>0$, $v_{0,y}=0$ no longer corresponds to critical points of dispersion $\epsilon_{\bf k}$ on the Fermi surface. 

It is worth noting that Berry curvature breaks time-reversal symmetry. Therefore, if the order of the two pulses is reversed, i.e., if $t_1>t_2$, the excess particle number in region $\Sigma$ would be different. It is meaningful to measure the average of these two different cases ($t_1<t_2$ and $t_1>t_2$), which is given by
\begin{align}
    \overline N=&\xi_1\xi_2\sum_{\bf v_0=0}\text{sgn}\det\left[\frac{\partial(v_{0,x},v_{0,y})}{\partial(k_x,k_y)}\right]\Bigg|_{\bf r=0}\nonumber\\
    &+\frac12\xi_1\xi_2\int d^2{\bf k}\Big\{ \Omega_{\bf k}[\theta(v_{0,x})\nabla_{\bf k } \theta(v_{0,y})\nonumber\\
    &\qquad-\theta(v_{0,y})\nabla_{\bf k } \theta(v_{0,x})]\cdot(\nabla_{\bf r} f_0)\Big|_{\bf r=0}\Big\}.
\end{align}
It is antisymmetric with respect to the exchange of $v_{0,x}$ and $v_{0,y}$.

In conclusion, due to the combined effect of Berry curvature and spatial inhomogeneity, anomalous velocities exist at equilibrium, which in turn affects nonlinear transport. The resultant nonlinear transport contains a quantized part and a non-quantized part. Neither of these two parts are solely determined by the Fermi sea topology. 
If the two pulses intersect at a local extremum, i.e., $\nabla_{\bf r}U_r\big|_{\bf r=0}=0$, the nonlinear transport is reduced to the standard case $N=\xi_1\xi_2\chi_F$.

The breakdown of the quantization can be observed in ultracold atoms. Topological bands with zero net magnetic flux can be readily implemented with current technology \cite{Cooper12,Jotzu14,Sun18}. By adjusting the filling factor of the optical lattice, one can realize a Fermi surface with nontrivial Berry curvature. The experimental protocol for nonlinear transport in ultracold atoms was analyzed in detail in Ref. \cite{Yang22QNL}. The voltage pulses can be simulated with far-detuned optical pulses and the excess particle number in region $\Sigma$ can be measured to single-atom precision with quantum gas microscope \cite{QGMRev}. When the two optical pulses intersect at the local extrema of the optical trap, $N$ is quantized in the limit $t_{31}\to0$. By varying the location of the intersection or the profile of the trap, one can observe the breakdown of quantization.

\section{Conclusions}

In this paper, we have shown that nontrival Berry curvatures at the Fermi surface do not affect quantized nonlinear transport in 2D ballistic metals as long as the sample is spatially homogeneous. This is because the nonlinear transport probes the properties of the metal at equilibrium where the anomalous velocities are absent. However, anomalous velocities enter when there exists a potential slowly varying in space. The combined effect of Berry curvature and the spatial inhomogeneity leads to the breakdown of the quantization, which can be observed in ultracold atoms.

{Another question one may ask after studying the Berry curvature effect is whether external magnetic field leads to similar results. Berry curvature can be viewed as magnetic field in momentum space. However, there are certain asymmetries between real space and momentum space in this setup. Therefore, we expect external magnetic field to lead to different physics. For example, for free electrons in an external magnetic field, we will have partially filled Landau level at the Fermi level. The transport properties are dictated by this partially filled Landau level instead of the free electron Fermi surface. The effect of magnetic field on nonlinear conductance is worth studying in the future.}

\begin{acknowledgments}
    We thank Hui Zhai for helpful discussions. This work is funeded by the Basic Research Program of Jiangsu (Grant No. BK20251338).
\end{acknowledgments}

\bibliography{references}

\begin{thebibliography}{33}%
\makeatletter
\providecommand \@ifxundefined [1]{%
 \@ifx{#1\undefined}
}%
\providecommand \@ifnum [1]{%
 \ifnum #1\expandafter \@firstoftwo
 \else \expandafter \@secondoftwo
 \fi
}%
\providecommand \@ifx [1]{%
 \ifx #1\expandafter \@firstoftwo
 \else \expandafter \@secondoftwo
 \fi
}%
\providecommand \natexlab [1]{#1}%
\providecommand \enquote  [1]{``#1''}%
\providecommand \bibnamefont  [1]{#1}%
\providecommand \bibfnamefont [1]{#1}%
\providecommand \citenamefont [1]{#1}%
\providecommand \href@noop [0]{\@secondoftwo}%
\providecommand \href [0]{\begingroup \@sanitize@url \@href}%
\providecommand \@href[1]{\@@startlink{#1}\@@href}%
\providecommand \@@href[1]{\endgroup#1\@@endlink}%
\providecommand \@sanitize@url [0]{\catcode `\\12\catcode `\$12\catcode `\&12\catcode `\#12\catcode `\^12\catcode `\_12\catcode `\%12\relax}%
\providecommand \@@startlink[1]{}%
\providecommand \@@endlink[0]{}%
\providecommand \url  [0]{\begingroup\@sanitize@url \@url }%
\providecommand \@url [1]{\endgroup\@href {#1}{\urlprefix }}%
\providecommand \urlprefix  [0]{URL }%
\providecommand \Eprint [0]{\href }%
\providecommand \doibase [0]{https://doi.org/}%
\providecommand \selectlanguage [0]{\@gobble}%
\providecommand \bibinfo  [0]{\@secondoftwo}%
\providecommand \bibfield  [0]{\@secondoftwo}%
\providecommand \translation [1]{[#1]}%
\providecommand \BibitemOpen [0]{}%
\providecommand \bibitemStop [0]{}%
\providecommand \bibitemNoStop [0]{.\EOS\space}%
\providecommand \EOS [0]{\spacefactor3000\relax}%
\providecommand \BibitemShut  [1]{\csname bibitem#1\endcsname}%
\let\auto@bib@innerbib\@empty
%</preamble>
\bibitem [{\citenamefont {Klitzing}\ \emph {et~al.}(1980)\citenamefont {Klitzing}, \citenamefont {Dorda},\ and\ \citenamefont {Pepper}}]{Klitzing80}%
  \BibitemOpen
  \bibfield  {author} {\bibinfo {author} {\bibfnamefont {K.~v.}\ \bibnamefont {Klitzing}}, \bibinfo {author} {\bibfnamefont {G.}~\bibnamefont {Dorda}},\ and\ \bibinfo {author} {\bibfnamefont {M.}~\bibnamefont {Pepper}},\ }\bibfield  {title} {\bibinfo {title} {{New Method for High-Accuracy Determination of the Fine-Structure Constant Based on Quantized Hall Resistance}},\ }\href {https://doi.org/10.1103/PhysRevLett.45.494} {\bibfield  {journal} {\bibinfo  {journal} {Phys. Rev. Lett.}\ }\textbf {\bibinfo {volume} {45}},\ \bibinfo {pages} {494} (\bibinfo {year} {1980})}\BibitemShut {NoStop}%
\bibitem [{\citenamefont {Thouless}\ \emph {et~al.}(1982)\citenamefont {Thouless}, \citenamefont {Kohmoto}, \citenamefont {Nightingale},\ and\ \citenamefont {den Nijs}}]{Thouless82}%
  \BibitemOpen
  \bibfield  {author} {\bibinfo {author} {\bibfnamefont {D.~J.}\ \bibnamefont {Thouless}}, \bibinfo {author} {\bibfnamefont {M.}~\bibnamefont {Kohmoto}}, \bibinfo {author} {\bibfnamefont {M.~P.}\ \bibnamefont {Nightingale}},\ and\ \bibinfo {author} {\bibfnamefont {M.}~\bibnamefont {den Nijs}},\ }\bibfield  {title} {\bibinfo {title} {{Quantized Hall Conductance in a Two-Dimensional Periodic Potential}},\ }\href {https://doi.org/10.1103/PhysRevLett.49.405} {\bibfield  {journal} {\bibinfo  {journal} {Phys. Rev. Lett.}\ }\textbf {\bibinfo {volume} {49}},\ \bibinfo {pages} {405} (\bibinfo {year} {1982})}\BibitemShut {NoStop}%
\bibitem [{\citenamefont {Kane}\ and\ \citenamefont {Mele}(2005)}]{Kane05b}%
  \BibitemOpen
  \bibfield  {author} {\bibinfo {author} {\bibfnamefont {C.~L.}\ \bibnamefont {Kane}}\ and\ \bibinfo {author} {\bibfnamefont {E.~J.}\ \bibnamefont {Mele}},\ }\bibfield  {title} {\bibinfo {title} {{Quantum Spin Hall Effect in Graphene}},\ }\href@noop {} {\bibfield  {journal} {\bibinfo  {journal} {Phys. Rev. Lett.}\ }\textbf {\bibinfo {volume} {95}},\ \bibinfo {pages} {226801} (\bibinfo {year} {2005})}\BibitemShut {NoStop}%
\bibitem [{\citenamefont {Bernevig}\ and\ \citenamefont {Zhang}(2006)}]{Bernevig06a}%
  \BibitemOpen
  \bibfield  {author} {\bibinfo {author} {\bibfnamefont {B.~A.}\ \bibnamefont {Bernevig}}\ and\ \bibinfo {author} {\bibfnamefont {S.~C.}\ \bibnamefont {Zhang}},\ }\bibfield  {title} {\bibinfo {title} {{Quantum Spin Hall Effect}},\ }\href@noop {} {\bibfield  {journal} {\bibinfo  {journal} {Phys. Rev. Lett.}\ }\textbf {\bibinfo {volume} {96}},\ \bibinfo {pages} {106802} (\bibinfo {year} {2006})}\BibitemShut {NoStop}%
\bibitem [{\citenamefont {Koenig}\ \emph {et~al.}(2007)\citenamefont {Koenig}, \citenamefont {Wiedmann}, \citenamefont {Bruene}, \citenamefont {Roth}, \citenamefont {Buhmann}, \citenamefont {Molenkamp}, \citenamefont {Qi},\ and\ \citenamefont {Zhang}}]{Konig07}%
  \BibitemOpen
  \bibfield  {author} {\bibinfo {author} {\bibfnamefont {M.}~\bibnamefont {Koenig}}, \bibinfo {author} {\bibfnamefont {S.}~\bibnamefont {Wiedmann}}, \bibinfo {author} {\bibfnamefont {C.}~\bibnamefont {Bruene}}, \bibinfo {author} {\bibfnamefont {A.}~\bibnamefont {Roth}}, \bibinfo {author} {\bibfnamefont {H.}~\bibnamefont {Buhmann}}, \bibinfo {author} {\bibfnamefont {L.~W.}\ \bibnamefont {Molenkamp}}, \bibinfo {author} {\bibfnamefont {X.-L.}\ \bibnamefont {Qi}},\ and\ \bibinfo {author} {\bibfnamefont {S.-C.}\ \bibnamefont {Zhang}},\ }\bibfield  {title} {\bibinfo {title} {Quantum spin hall insulator state in hgte quantum wells},\ }\href {https://doi.org/10.1126/science.1148047} {\bibfield  {journal} {\bibinfo  {journal} {Science}\ }\textbf {\bibinfo {volume} {318}},\ \bibinfo {pages} {766} (\bibinfo {year} {2007})}\BibitemShut {NoStop}%
\bibitem [{\citenamefont {Landauer}(1957)}]{Landauer57}%
  \BibitemOpen
  \bibfield  {author} {\bibinfo {author} {\bibfnamefont {R.}~\bibnamefont {Landauer}},\ }\bibfield  {title} {\bibinfo {title} {Spatial variation of currents and fields due to localized scatterers in metallic conduction},\ }\href {https://doi.org/10.1147/rd.13.0223} {\bibfield  {journal} {\bibinfo  {journal} {IBM J. Res. Dev.}\ }\textbf {\bibinfo {volume} {1}},\ \bibinfo {pages} {223} (\bibinfo {year} {1957})}\BibitemShut {NoStop}%
\bibitem [{\citenamefont {van Wees}\ \emph {et~al.}(1988)\citenamefont {van Wees}, \citenamefont {van Houten}, \citenamefont {Beenakker}, \citenamefont {Williamson}, \citenamefont {Kouwenhoven}, \citenamefont {van~der Marel},\ and\ \citenamefont {Foxon}}]{vanWees88}%
  \BibitemOpen
  \bibfield  {author} {\bibinfo {author} {\bibfnamefont {B.~J.}\ \bibnamefont {van Wees}}, \bibinfo {author} {\bibfnamefont {H.}~\bibnamefont {van Houten}}, \bibinfo {author} {\bibfnamefont {C.~W.~J.}\ \bibnamefont {Beenakker}}, \bibinfo {author} {\bibfnamefont {J.~G.}\ \bibnamefont {Williamson}}, \bibinfo {author} {\bibfnamefont {L.~P.}\ \bibnamefont {Kouwenhoven}}, \bibinfo {author} {\bibfnamefont {D.}~\bibnamefont {van~der Marel}},\ and\ \bibinfo {author} {\bibfnamefont {C.~T.}\ \bibnamefont {Foxon}},\ }\bibfield  {title} {\bibinfo {title} {Quantized conductance of point contacts in a two-dimensional electron gas},\ }\href {https://doi.org/10.1103/PhysRevLett.60.848} {\bibfield  {journal} {\bibinfo  {journal} {Phys. Rev. Lett.}\ }\textbf {\bibinfo {volume} {60}},\ \bibinfo {pages} {848} (\bibinfo {year} {1988})}\BibitemShut {NoStop}%
\bibitem [{\citenamefont {Wharam}\ \emph {et~al.}(1988)\citenamefont {Wharam}, \citenamefont {Thornton}, \citenamefont {Newbury}, \citenamefont {Pepper}, \citenamefont {Ahmed}, \citenamefont {Frost}, \citenamefont {Hasko}, \citenamefont {Peacock}, \citenamefont {Ritchie},\ and\ \citenamefont {Jones}}]{Wharam88}%
  \BibitemOpen
  \bibfield  {author} {\bibinfo {author} {\bibfnamefont {D.~A.}\ \bibnamefont {Wharam}}, \bibinfo {author} {\bibfnamefont {T.~J.}\ \bibnamefont {Thornton}}, \bibinfo {author} {\bibfnamefont {R.}~\bibnamefont {Newbury}}, \bibinfo {author} {\bibfnamefont {M.}~\bibnamefont {Pepper}}, \bibinfo {author} {\bibfnamefont {H.}~\bibnamefont {Ahmed}}, \bibinfo {author} {\bibfnamefont {J.~E.~F.}\ \bibnamefont {Frost}}, \bibinfo {author} {\bibfnamefont {D.~G.}\ \bibnamefont {Hasko}}, \bibinfo {author} {\bibfnamefont {D.~C.}\ \bibnamefont {Peacock}}, \bibinfo {author} {\bibfnamefont {D.~A.}\ \bibnamefont {Ritchie}},\ and\ \bibinfo {author} {\bibfnamefont {G.~A.~C.}\ \bibnamefont {Jones}},\ }\bibfield  {title} {\bibinfo {title} {One-dimensional transport and the quantisation of the ballistic resistance},\ }\href {https://doi.org/10.1088/0022-3719/21/8/002} {\bibfield  {journal} {\bibinfo  {journal} {Journal of Physics C: Solid State Physics}\ }\textbf {\bibinfo {volume} {21}},\ \bibinfo {pages} {L209} (\bibinfo {year}
  {1988})}\BibitemShut {NoStop}%
\bibitem [{\citenamefont {Honda}\ \emph {et~al.}(1995)\citenamefont {Honda}, \citenamefont {Tarucha}, \citenamefont {Saku},\ and\ \citenamefont {Yasuhiro~Tokura}}]{Honda95}%
  \BibitemOpen
  \bibfield  {author} {\bibinfo {author} {\bibfnamefont {T.}~\bibnamefont {Honda}}, \bibinfo {author} {\bibfnamefont {S.}~\bibnamefont {Tarucha}}, \bibinfo {author} {\bibfnamefont {T.}~\bibnamefont {Saku}},\ and\ \bibinfo {author} {\bibfnamefont {Y.~T.}\ \bibnamefont {Yasuhiro~Tokura}},\ }\bibfield  {title} {\bibinfo {title} {Quantized conductance observed in quantum wires 2 to 10 $\mu m$ long},\ }\href {https://doi.org/10.1143/JJAP.34.L72} {\bibfield  {journal} {\bibinfo  {journal} {Japanese Journal of Applied Physics}\ }\textbf {\bibinfo {volume} {34}},\ \bibinfo {pages} {L72} (\bibinfo {year} {1995})}\BibitemShut {NoStop}%
\bibitem [{\citenamefont {van Weperen}\ \emph {et~al.}(2013)\citenamefont {van Weperen}, \citenamefont {Plissard}, \citenamefont {Bakkers}, \citenamefont {Frolov},\ and\ \citenamefont {Kouwenhoven}}]{vanWeperen13}%
  \BibitemOpen
  \bibfield  {author} {\bibinfo {author} {\bibfnamefont {I.}~\bibnamefont {van Weperen}}, \bibinfo {author} {\bibfnamefont {S.~R.}\ \bibnamefont {Plissard}}, \bibinfo {author} {\bibfnamefont {E.~P. A.~M.}\ \bibnamefont {Bakkers}}, \bibinfo {author} {\bibfnamefont {S.~M.}\ \bibnamefont {Frolov}},\ and\ \bibinfo {author} {\bibfnamefont {L.~P.}\ \bibnamefont {Kouwenhoven}},\ }\bibfield  {title} {\bibinfo {title} {Quantized conductance in an insb nanowire},\ }\href {https://doi.org/10.1021/nl3035256} {\bibfield  {journal} {\bibinfo  {journal} {Nano Letters}\ }\textbf {\bibinfo {volume} {13}},\ \bibinfo {pages} {387} (\bibinfo {year} {2013})}\BibitemShut {NoStop}%
\bibitem [{\citenamefont {Frank}\ \emph {et~al.}(1998)\citenamefont {Frank}, \citenamefont {Poncharal}, \citenamefont {Wang},\ and\ \citenamefont {de~Heer}}]{Frank98}%
  \BibitemOpen
  \bibfield  {author} {\bibinfo {author} {\bibfnamefont {S.}~\bibnamefont {Frank}}, \bibinfo {author} {\bibfnamefont {P.}~\bibnamefont {Poncharal}}, \bibinfo {author} {\bibfnamefont {Z.~L.}\ \bibnamefont {Wang}},\ and\ \bibinfo {author} {\bibfnamefont {W.~A.}\ \bibnamefont {de~Heer}},\ }\bibfield  {title} {\bibinfo {title} {Carbon nanotube quantum resistors},\ }\href {https://doi.org/10.1126/science.280.5370.1744} {\bibfield  {journal} {\bibinfo  {journal} {Science}\ }\textbf {\bibinfo {volume} {280}},\ \bibinfo {pages} {1744} (\bibinfo {year} {1998})}\BibitemShut {NoStop}%
\bibitem [{\citenamefont {Kane}(2022)}]{Kane22}%
  \BibitemOpen
  \bibfield  {author} {\bibinfo {author} {\bibfnamefont {C.~L.}\ \bibnamefont {Kane}},\ }\bibfield  {title} {\bibinfo {title} {{Quantized Nonlinear Conductance in Ballistic Metals}},\ }\href {https://doi.org/10.1103/PhysRevLett.128.076801} {\bibfield  {journal} {\bibinfo  {journal} {Phys. Rev. Lett.}\ }\textbf {\bibinfo {volume} {128}},\ \bibinfo {pages} {076801} (\bibinfo {year} {2022})}\BibitemShut {NoStop}%
\bibitem [{\citenamefont {Krinner}\ \emph {et~al.}(2015)\citenamefont {Krinner}, \citenamefont {Stadler}, \citenamefont {Husmann}, \citenamefont {Brantut},\ and\ \citenamefont {Esslinger}}]{Krinner15}%
  \BibitemOpen
  \bibfield  {author} {\bibinfo {author} {\bibfnamefont {S.}~\bibnamefont {Krinner}}, \bibinfo {author} {\bibfnamefont {D.}~\bibnamefont {Stadler}}, \bibinfo {author} {\bibfnamefont {D.}~\bibnamefont {Husmann}}, \bibinfo {author} {\bibfnamefont {J.-P.}\ \bibnamefont {Brantut}},\ and\ \bibinfo {author} {\bibfnamefont {T.}~\bibnamefont {Esslinger}},\ }\bibfield  {title} {\bibinfo {title} {Observation of quantized conductance in neutral matter},\ }\href@noop {} {\bibfield  {journal} {\bibinfo  {journal} {Nature}\ }\textbf {\bibinfo {volume} {517}},\ \bibinfo {pages} {64} (\bibinfo {year} {2015})}\BibitemShut {NoStop}%
\bibitem [{\citenamefont {Krinner}\ \emph {et~al.}(2017)\citenamefont {Krinner}, \citenamefont {Esslinger},\ and\ \citenamefont {Brantut}}]{Krinner17}%
  \BibitemOpen
  \bibfield  {author} {\bibinfo {author} {\bibfnamefont {S.}~\bibnamefont {Krinner}}, \bibinfo {author} {\bibfnamefont {T.}~\bibnamefont {Esslinger}},\ and\ \bibinfo {author} {\bibfnamefont {J.-P.}\ \bibnamefont {Brantut}},\ }\bibfield  {title} {\bibinfo {title} {Two-terminal transport measurements with cold atoms},\ }\href {https://doi.org/10.1088/1361-648X/aa74a1} {\bibfield  {journal} {\bibinfo  {journal} {Journal of Physics: Condensed Matter}\ }\textbf {\bibinfo {volume} {29}},\ \bibinfo {pages} {343003} (\bibinfo {year} {2017})}\BibitemShut {NoStop}%
\bibitem [{\citenamefont {Lebrat}\ \emph {et~al.}(2019)\citenamefont {Lebrat}, \citenamefont {H\"ausler}, \citenamefont {Fabritius}, \citenamefont {Husmann}, \citenamefont {Corman},\ and\ \citenamefont {Esslinger}}]{Lebrat19}%
  \BibitemOpen
  \bibfield  {author} {\bibinfo {author} {\bibfnamefont {M.}~\bibnamefont {Lebrat}}, \bibinfo {author} {\bibfnamefont {S.}~\bibnamefont {H\"ausler}}, \bibinfo {author} {\bibfnamefont {P.}~\bibnamefont {Fabritius}}, \bibinfo {author} {\bibfnamefont {D.}~\bibnamefont {Husmann}}, \bibinfo {author} {\bibfnamefont {L.}~\bibnamefont {Corman}},\ and\ \bibinfo {author} {\bibfnamefont {T.}~\bibnamefont {Esslinger}},\ }\bibfield  {title} {\bibinfo {title} {Quantized conductance through a spin-selective atomic point contact},\ }\href {https://doi.org/10.1103/PhysRevLett.123.193605} {\bibfield  {journal} {\bibinfo  {journal} {Phys. Rev. Lett.}\ }\textbf {\bibinfo {volume} {123}},\ \bibinfo {pages} {193605} (\bibinfo {year} {2019})}\BibitemShut {NoStop}%
\bibitem [{\citenamefont {Yang}\ and\ \citenamefont {Zhai}(2022)}]{Yang22QNL}%
  \BibitemOpen
  \bibfield  {author} {\bibinfo {author} {\bibfnamefont {F.}~\bibnamefont {Yang}}\ and\ \bibinfo {author} {\bibfnamefont {H.}~\bibnamefont {Zhai}},\ }\bibfield  {title} {\bibinfo {title} {Quantized {N}onlinear {T}ransport with {U}ltracold {A}toms},\ }\href {https://doi.org/10.22331/q-2022-11-10-857} {\bibfield  {journal} {\bibinfo  {journal} {{Quantum}}\ }\textbf {\bibinfo {volume} {6}},\ \bibinfo {pages} {857} (\bibinfo {year} {2022})}\BibitemShut {NoStop}%
\bibitem [{\citenamefont {Zhang}(2023)}]{PFZhang23}%
  \BibitemOpen
  \bibfield  {author} {\bibinfo {author} {\bibfnamefont {P.}~\bibnamefont {Zhang}},\ }\bibfield  {title} {\bibinfo {title} {Quantized topological response in trapped quantum gases},\ }\href {https://doi.org/10.1103/PhysRevA.107.L031305} {\bibfield  {journal} {\bibinfo  {journal} {Phys. Rev. A}\ }\textbf {\bibinfo {volume} {107}},\ \bibinfo {pages} {L031305} (\bibinfo {year} {2023})}\BibitemShut {NoStop}%
\bibitem [{\citenamefont {Tam}\ and\ \citenamefont {Kane}(2023)}]{Tam23}%
  \BibitemOpen
  \bibfield  {author} {\bibinfo {author} {\bibfnamefont {P.~M.}\ \bibnamefont {Tam}}\ and\ \bibinfo {author} {\bibfnamefont {C.~L.}\ \bibnamefont {Kane}},\ }\bibfield  {title} {\bibinfo {title} {{Probing Fermi Sea Topology by Andreev State Transport}},\ }\href {https://doi.org/10.1103/PhysRevLett.130.096301} {\bibfield  {journal} {\bibinfo  {journal} {Phys. Rev. Lett.}\ }\textbf {\bibinfo {volume} {130}},\ \bibinfo {pages} {096301} (\bibinfo {year} {2023})}\BibitemShut {NoStop}%
\bibitem [{\citenamefont {Tam}\ \emph {et~al.}(2023)\citenamefont {Tam}, \citenamefont {De~Beule},\ and\ \citenamefont {Kane}}]{tam2023topological}%
  \BibitemOpen
  \bibfield  {author} {\bibinfo {author} {\bibfnamefont {P.~M.}\ \bibnamefont {Tam}}, \bibinfo {author} {\bibfnamefont {C.}~\bibnamefont {De~Beule}},\ and\ \bibinfo {author} {\bibfnamefont {C.~L.}\ \bibnamefont {Kane}},\ }\bibfield  {title} {\bibinfo {title} {Topological {A}ndreev rectification},\ }\href {https://doi.org/10.1103/PhysRevB.107.245422} {\bibfield  {journal} {\bibinfo  {journal} {Phys. Rev. B}\ }\textbf {\bibinfo {volume} {107}},\ \bibinfo {pages} {245422} (\bibinfo {year} {2023})}\BibitemShut {NoStop}%
\bibitem [{\citenamefont {Tam}\ \emph {et~al.}(2022)\citenamefont {Tam}, \citenamefont {Claassen},\ and\ \citenamefont {Kane}}]{Tam22}%
  \BibitemOpen
  \bibfield  {author} {\bibinfo {author} {\bibfnamefont {P.~M.}\ \bibnamefont {Tam}}, \bibinfo {author} {\bibfnamefont {M.}~\bibnamefont {Claassen}},\ and\ \bibinfo {author} {\bibfnamefont {C.~L.}\ \bibnamefont {Kane}},\ }\bibfield  {title} {\bibinfo {title} {Topological multipartite entanglement in a {F}ermi liquid},\ }\href {https://doi.org/10.1103/PhysRevX.12.031022} {\bibfield  {journal} {\bibinfo  {journal} {Phys. Rev. X}\ }\textbf {\bibinfo {volume} {12}},\ \bibinfo {pages} {031022} (\bibinfo {year} {2022})}\BibitemShut {NoStop}%
\bibitem [{\citenamefont {Tam}\ and\ \citenamefont {Kane}(2024)}]{Tam24}%
  \BibitemOpen
  \bibfield  {author} {\bibinfo {author} {\bibfnamefont {P.~M.}\ \bibnamefont {Tam}}\ and\ \bibinfo {author} {\bibfnamefont {C.~L.}\ \bibnamefont {Kane}},\ }\bibfield  {title} {\bibinfo {title} {Topological density correlations in a fermi gas},\ }\href {https://doi.org/10.1103/PhysRevB.109.035413} {\bibfield  {journal} {\bibinfo  {journal} {Phys. Rev. B}\ }\textbf {\bibinfo {volume} {109}},\ \bibinfo {pages} {035413} (\bibinfo {year} {2024})}\BibitemShut {NoStop}%
\bibitem [{\citenamefont {Yang}\ \emph {et~al.}(2023)\citenamefont {Yang}, \citenamefont {Li},\ and\ \citenamefont {Li}}]{EulerChern}%
  \BibitemOpen
  \bibfield  {author} {\bibinfo {author} {\bibfnamefont {F.}~\bibnamefont {Yang}}, \bibinfo {author} {\bibfnamefont {X.}~\bibnamefont {Li}},\ and\ \bibinfo {author} {\bibfnamefont {C.}~\bibnamefont {Li}},\ }\bibfield  {title} {\bibinfo {title} {Euler-chern correspondence via topological superconductivity},\ }\href {https://doi.org/10.1103/PhysRevResearch.5.033073} {\bibfield  {journal} {\bibinfo  {journal} {Phys. Rev. Res.}\ }\textbf {\bibinfo {volume} {5}},\ \bibinfo {pages} {033073} (\bibinfo {year} {2023})}\BibitemShut {NoStop}%
\bibitem [{\citenamefont {Jia}(2025)}]{Jia24}%
  \BibitemOpen
  \bibfield  {author} {\bibinfo {author} {\bibfnamefont {W.}~\bibnamefont {Jia}},\ }\bibfield  {title} {\bibinfo {title} {Generic reduction theory for fermi sea topology in metallic systems},\ }\href {https://doi.org/10.1103/PhysRevB.111.155115} {\bibfield  {journal} {\bibinfo  {journal} {Phys. Rev. B}\ }\textbf {\bibinfo {volume} {111}},\ \bibinfo {pages} {155115} (\bibinfo {year} {2025})}\BibitemShut {NoStop}%
\bibitem [{\citenamefont {Haldane}(2004)}]{Haldane04}%
  \BibitemOpen
  \bibfield  {author} {\bibinfo {author} {\bibfnamefont {F.~D.~M.}\ \bibnamefont {Haldane}},\ }\bibfield  {title} {\bibinfo {title} {Berry curvature on the fermi surface: Anomalous hall effect as a topological fermi-liquid property},\ }\href {https://doi.org/10.1103/PhysRevLett.93.206602} {\bibfield  {journal} {\bibinfo  {journal} {Phys. Rev. Lett.}\ }\textbf {\bibinfo {volume} {93}},\ \bibinfo {pages} {206602} (\bibinfo {year} {2004})}\BibitemShut {NoStop}%
\bibitem [{\citenamefont {Sodemann}\ and\ \citenamefont {Fu}(2015)}]{Sodemann15}%
  \BibitemOpen
  \bibfield  {author} {\bibinfo {author} {\bibfnamefont {I.}~\bibnamefont {Sodemann}}\ and\ \bibinfo {author} {\bibfnamefont {L.}~\bibnamefont {Fu}},\ }\bibfield  {title} {\bibinfo {title} {Quantum nonlinear hall effect induced by berry curvature dipole in time-reversal invariant materials},\ }\href {https://doi.org/10.1103/PhysRevLett.115.216806} {\bibfield  {journal} {\bibinfo  {journal} {Phys. Rev. Lett.}\ }\textbf {\bibinfo {volume} {115}},\ \bibinfo {pages} {216806} (\bibinfo {year} {2015})}\BibitemShut {NoStop}%
\bibitem [{\citenamefont {Chen}\ and\ \citenamefont {Zheng}(2022)}]{XLChen22}%
  \BibitemOpen
  \bibfield  {author} {\bibinfo {author} {\bibfnamefont {X.-L.}\ \bibnamefont {Chen}}\ and\ \bibinfo {author} {\bibfnamefont {W.}~\bibnamefont {Zheng}},\ }\bibfield  {title} {\bibinfo {title} {Nonlinear hall response in the driving dynamics of ultracold atoms in optical lattices},\ }\href {https://doi.org/10.1103/PhysRevA.105.063312} {\bibfield  {journal} {\bibinfo  {journal} {Phys. Rev. A}\ }\textbf {\bibinfo {volume} {105}},\ \bibinfo {pages} {063312} (\bibinfo {year} {2022})}\BibitemShut {NoStop}%
\bibitem [{\citenamefont {Xiao}\ \emph {et~al.}(2010)\citenamefont {Xiao}, \citenamefont {Chang},\ and\ \citenamefont {Niu}}]{Xiao10}%
  \BibitemOpen
  \bibfield  {author} {\bibinfo {author} {\bibfnamefont {D.}~\bibnamefont {Xiao}}, \bibinfo {author} {\bibfnamefont {M.-C.}\ \bibnamefont {Chang}},\ and\ \bibinfo {author} {\bibfnamefont {Q.}~\bibnamefont {Niu}},\ }\bibfield  {title} {\bibinfo {title} {Berry phase effects on electronic properties},\ }\href {https://doi.org/10.1103/RevModPhys.82.1959} {\bibfield  {journal} {\bibinfo  {journal} {Rev. Mod. Phys.}\ }\textbf {\bibinfo {volume} {82}},\ \bibinfo {pages} {1959} (\bibinfo {year} {2010})}\BibitemShut {NoStop}%
\bibitem [{\citenamefont {Courant}\ and\ \citenamefont {Hilbert}(1989)}]{MathPhys}%
  \BibitemOpen
  \bibfield  {author} {\bibinfo {author} {\bibfnamefont {R.}~\bibnamefont {Courant}}\ and\ \bibinfo {author} {\bibfnamefont {D.}~\bibnamefont {Hilbert}},\ }\href@noop {} {\emph {\bibinfo {title} {Methods of Mathematical Physics}}}\ (\bibinfo  {publisher} {John Weley and Sons},\ \bibinfo {year} {1989})\BibitemShut {NoStop}%
\bibitem [{\citenamefont {Milnor}(1997)}]{TopoDiff}%
  \BibitemOpen
  \bibfield  {author} {\bibinfo {author} {\bibfnamefont {J.}~\bibnamefont {Milnor}},\ }\href {https://press.princeton.edu/books/paperback/9780691048338/topology-from-the-differentiable-viewpoint} {\emph {\bibinfo {title} {Topology from the Differentiable Viewpoint}}}\ (\bibinfo  {publisher} {Princeton University Press},\ \bibinfo {year} {1997})\BibitemShut {NoStop}%
\bibitem [{\citenamefont {Cooper}\ and\ \citenamefont {Moessner}(2012)}]{Cooper12}%
  \BibitemOpen
  \bibfield  {author} {\bibinfo {author} {\bibfnamefont {N.~R.}\ \bibnamefont {Cooper}}\ and\ \bibinfo {author} {\bibfnamefont {R.}~\bibnamefont {Moessner}},\ }\bibfield  {title} {\bibinfo {title} {Designing topological bands in reciprocal space},\ }\href {https://doi.org/10.1103/PhysRevLett.109.215302} {\bibfield  {journal} {\bibinfo  {journal} {Phys. Rev. Lett.}\ }\textbf {\bibinfo {volume} {109}},\ \bibinfo {pages} {215302} (\bibinfo {year} {2012})}\BibitemShut {NoStop}%
\bibitem [{\citenamefont {Jotzu}\ \emph {et~al.}(2014)\citenamefont {Jotzu}, \citenamefont {Messer}, \citenamefont {Desbuquois}, \citenamefont {Lebrat}, \citenamefont {Uehlinger}, \citenamefont {Greif},\ and\ \citenamefont {Esslinger}}]{Jotzu14}%
  \BibitemOpen
  \bibfield  {author} {\bibinfo {author} {\bibfnamefont {G.}~\bibnamefont {Jotzu}}, \bibinfo {author} {\bibfnamefont {M.}~\bibnamefont {Messer}}, \bibinfo {author} {\bibfnamefont {R.}~\bibnamefont {Desbuquois}}, \bibinfo {author} {\bibfnamefont {M.}~\bibnamefont {Lebrat}}, \bibinfo {author} {\bibfnamefont {T.}~\bibnamefont {Uehlinger}}, \bibinfo {author} {\bibfnamefont {D.}~\bibnamefont {Greif}},\ and\ \bibinfo {author} {\bibfnamefont {T.}~\bibnamefont {Esslinger}},\ }\bibfield  {title} {\bibinfo {title} {Experimental realization of the topological haldane model with ultracold fermions},\ }\href@noop {} {\bibfield  {journal} {\bibinfo  {journal} {Nature}\ ,\ \bibinfo {pages} {237}} (\bibinfo {year} {2014})}\BibitemShut {NoStop}%
\bibitem [{\citenamefont {Sun}\ \emph {et~al.}(2018)\citenamefont {Sun}, \citenamefont {Wang}, \citenamefont {Xu}, \citenamefont {Yi}, \citenamefont {Zhang}, \citenamefont {Wu}, \citenamefont {Deng}, \citenamefont {Liu}, \citenamefont {Chen},\ and\ \citenamefont {Pan}}]{Sun18}%
  \BibitemOpen
  \bibfield  {author} {\bibinfo {author} {\bibfnamefont {W.}~\bibnamefont {Sun}}, \bibinfo {author} {\bibfnamefont {B.-Z.}\ \bibnamefont {Wang}}, \bibinfo {author} {\bibfnamefont {X.-T.}\ \bibnamefont {Xu}}, \bibinfo {author} {\bibfnamefont {C.-R.}\ \bibnamefont {Yi}}, \bibinfo {author} {\bibfnamefont {L.}~\bibnamefont {Zhang}}, \bibinfo {author} {\bibfnamefont {Z.}~\bibnamefont {Wu}}, \bibinfo {author} {\bibfnamefont {Y.}~\bibnamefont {Deng}}, \bibinfo {author} {\bibfnamefont {X.-J.}\ \bibnamefont {Liu}}, \bibinfo {author} {\bibfnamefont {S.}~\bibnamefont {Chen}},\ and\ \bibinfo {author} {\bibfnamefont {J.-W.}\ \bibnamefont {Pan}},\ }\bibfield  {title} {\bibinfo {title} {Highly controllable and robust 2d spin-orbit coupling for quantum gases},\ }\href {https://doi.org/10.1103/PhysRevLett.121.150401} {\bibfield  {journal} {\bibinfo  {journal} {Phys. Rev. Lett.}\ }\textbf {\bibinfo {volume} {121}},\ \bibinfo {pages} {150401} (\bibinfo {year} {2018})}\BibitemShut {NoStop}%
\bibitem [{\citenamefont {Gross}\ and\ \citenamefont {Bakr}(2021)}]{QGMRev}%
  \BibitemOpen
  \bibfield  {author} {\bibinfo {author} {\bibfnamefont {C.}~\bibnamefont {Gross}}\ and\ \bibinfo {author} {\bibfnamefont {W.}~\bibnamefont {Bakr}},\ }\bibfield  {title} {\bibinfo {title} {Quantum gas microscopy for single atom and spin detection},\ }\href {https://doi.org/doi.org/10.1038/s41567-021-01370-5} {\bibfield  {journal} {\bibinfo  {journal} {Nature Physics}\ }\textbf {\bibinfo {volume} {17}},\ \bibinfo {pages} {1316} (\bibinfo {year} {2021})}\BibitemShut {NoStop}%
\end{thebibliography}%

\end{document}